# THERMODYNAMICS OF KETONE + AMINE MIXTURES PART II. VOLUMETRIC AND SPEED OF SOUND DATA AT (293.15, 298.15 AND 303.15) K FOR 2-PROPANONE + DIPROPYLAMINE, + DIBUTYLAMINE OR + TRIETHYLAMINE SYSTEMS

IVÁN ALONSO, VÍCTOR ALONSO, ISMAEL MOZO, ISAÍAS GARCÍA DE LA FUENTE, JUAN ANTONIO GONZÁLEZ[*], AND JOSÉ CARLOS COBOS

G.E.T.E.F., Departamento de Física Aplicada, Facultad de Ciencias, Universidad de Valladolid, 47071 Valladolid, Spain,

*e-mail: jagl@termo.uva.es; Fax: +34-983-423136; Tel: +34-983-423757

**Abstract**

Densities, $\rho$, and speeds of sound, $u$, of 2-propanone + dipropylamine, + dibutylamine or + triethylamine systems have been measured at (293.15, 298.15 and 303.15) K and atmospheric pressure using a vibrating tube densimeter and sound analyser Anton Paar model DSA-5000. The $\rho$ and $u$ values were used to calculate excess molar volumes, $V^E$, and the excess functions at 298.15 K for the thermal expansion coefficient, $\alpha_P^E$, and for the isentropic compressibility, $\kappa_S^E$ at 298.15 K. $V^E$, $\kappa_S^E$ and $\alpha_P^E$ are positive magnitudes. When replacing dipropylamine by dibutylamine or triethylamine in the studied mixtures, the excess functions increase. This may be ascribed to the interactions between unlike molecules are more important in the former solutions. From the comparison with similar data obtained for 2-propanone + aniline, + $N$-methylaniline, or + pyridine systems, it is concluded that interactions between unlike molecules are stronger in mixtures containing aromatic amines. Free volume effects are present in solutions with dipropyl or dibutylamine as the $V^E$ curves are shifted towards higher mole fractions of 2-propanone.

## 1. Introduction

Amides, amino acids, peptides and their derivatives are of interest because they are simple models in biochemistry. *N*-methylformamide possesses the basic ($-CO$) and acidic ($-NH$) groups of the very common, in nature, peptide bond [1]. For example, proteins are polymers of amino acids linked to each other by peptide bonds. Consequently, the understanding of liquid mixtures involving the amide functional group is necessary as a first step to a better knowledge of complex molecules of biological interest [2]. So, the aqueous solution of dimethylformamide is a model solvent representing the environment of the interior of proteins. Amides have many practical applications. Dimethylformamide and *N*-methylpyrrolidone are used as highly selective extractants for the recovery of aromatic and saturated hydrocarbons from petroleum feedstocks [3], and $\varepsilon$-caprolactam is used for the production of nylon 6, which is a polycaprolactam formed by ring-opening polymerization. The study of alkanone + amine mixtures, which contain the carbonyl and amine groups in separate molecules, is then pertinent in order to gain insight into amide solutions. In this article, we report densities, speeds of sound and excess molar volumes at 293.15 K, 298.15 K and 303.15 K, and the excess functions at 298.15 K for the isobaric thermal expansion coefficients and the isentropic compressibility for the mixtures 2-propanone + dipropylamine, + dibutylamine or + triethylamine. In the first work of this series we have provided similar data for 2-propanone + aniline, + *N*-methylaniline, or + pyridine [4].

## 2. Experimental

### 2.1 Materials

2-Propanone ($\geq$ 0.995) was from Sigma Aldrich; dipropylamine ($\geq$ 0.99) and dibutylmaine ($\geq$ 0.995) were from Aldrich and triethylamine ($\geq$ 0.995) was form Fluka and used without further purification (purities expressed in mass fraction). The $\rho$ and $u$ values of the pure liquids are in good agreement with those from the literature (Table 1).

### 2.2 Apparatus and procedure

Binary mixtures were prepared by mass in small vessels of about 10 cm$^3$. Caution was taken to prevent evaporation, and the error in the final mole fraction is estimated to be less than $\pm$ 0.0001. Conversion to molar quantities was based on the relative atomic mass table of 2006 issued by IUPAC [5].

The densities and speeds of sound of both pure liquids and of the mixtures were measured using a vibrating-tube densimeter and sound analyser, Anton Paar model DSA-5000, automatically thermostated within $\pm$ 0.01 K. The calibration of the apparatus was carried out with deionised double-distilled water, heptane, octane, isooctane, cyclohexane and benzene, using $\rho$ values from the literature [6-8]. The accuracy for the $\rho$ and $u$ measurements are $\pm 1 \cdot 10^{-2}$ kg·m$^{-3}$ and $\pm 0.1$ m s$^{-1}$, respectively, and the corresponding precisions are $\pm 1 \cdot 10^{-3}$ kg m$^{-3}$ and $\pm 0.01$ m s$^{-1}$. The experimental technique was checked by determining $V^E$ and $u$ of the standard mixtures: (cyclohexane + benzene) at the temperatures (293.15, 298.15 and 303.15) K and 2-ethoxyethanol + heptane at 298.15 K. Our results agree well with published values [9-12]. The accuracy in $V^E$ is believed to be less than $\pm(0.01|V^E_{max}|+0.005)$ cm$^3$ mol$^{-1}$, where $|V^E_{max}|$ denotes the maximum experimental value of the excess molar volume with respect to the mole fraction.

### 3. Equations

The thermodynamic properties for which values are derived most directly from the experimental measurements are the density, $\rho$, the molar volume, $V$, the coefficient of thermal expansion, $\alpha_P = -\dfrac{1}{\rho}\left(\dfrac{\partial \rho}{\partial T}\right)_P$ and the isentropic compressibility, $\kappa_S$. In this work, $\alpha_P$ values were obtained from a linear dependence of $\rho$ with $T$. Assuming that the absorption of the acoustic wave is negligible, $\kappa_s$ can be calculated using the Newton-Laplace's equation:

$$\kappa_S = \frac{1}{\rho u^2} \qquad (1)$$

For an ideal mixture at the same temperature and pressure than the system under study, the values $F^{id}$ of the thermodynamic property, $F$, are calculated using the equations [9,13]:

$$F^{id} = x_1 F_1 + x_2 F_2 \qquad (F = V; C_P) \qquad (2)$$

and

$$F^{id} = \phi_1 F_1 + \phi_2 F_2 \qquad (F = \alpha_P; \kappa_T) \qquad (3)$$

where $C_p$ is the isobaric heat capacity, $\phi_i = \dfrac{x_i V_i}{V^{id}}$ the volume fraction, $\kappa_T$, the isothermal compressibility, and $F_i$, the $F$ value of component i, respectively. For $\kappa_S$ the ideal values are calculated according to [13]:

$$\kappa_S^{id} = \kappa_T^{id} - \frac{TV^{id}\alpha_P^{id2}}{C_P^{id}} \tag{4}$$

In this work, we have determined the excess functions:

$$F^E = F - F^{id} \tag{5}$$

for $F = V^E$, $\kappa_S$ and $\alpha_P$

## 4. Results and Discussion

Table 2 lists values of densities, calculated $V^E$ and of $u$ vs. $x_1$, the mole fraction of the 2-propanone. Table 3 contains the derived quantities $\kappa_S^E$ and $\alpha_P^E$. The data were fitted by unweighted least-squares polynomial regression to the equation:

$$F^E = x_1(1-x_1)\sum_{i=0}^{k-1} A_i(2x_1-1)^i \tag{6}$$

where $F$ stands for the properties cited above. The number of coefficients $k$ used in eq. (6) for each mixture was determined by applying an F-test [14] at the 99.5 % confidence level. Table 4 lists the parameters $A_i$ obtained in the regression, together with the standard deviations $\sigma$, defined by:

$$\sigma\left(F^E\right) = \left[\frac{1}{N-k}\sum\left(F_{cal}^E - F_{exp}^E\right)^2\right]^{1/2} \tag{7}$$

where $N$ is the number of direct experimental values. Results on $V^E$ and $\kappa_S^E$ are shown graphically in Figs 1 and 2. No data have been encountered in the literature for comparison.

Hereafter, we are referring to values of the excess molar properties at equimolar composition and 298.15 K.

Mixtures of 2-propanone with a given alkane are characterized by strong dipolar interactions between the ketone molecules, which leads to miscibility gaps at temperatures near to 298.15. For the heptane system, the upper critical solution temperature is 245.22 K [15], and the $H^E$ and $V^E$ values are 1704 J•mol$^{-1}$ [16] and = 1.130 cm$^3$•mol$^{-1}$ [17], respectively. The large positive $V^E$ value indicates that the interactional contribution to this excess function, due to the disruption of the ketone-ketone interactions upon mixing, is much more important than those related to effects which contribute negatively to $V^E$ (structural effects arising from interstitial accommodation of one component into the other and free volume effects). Dipropylamine and dibutylamine are secondary amines, and are weakly self-associated [18]. Accordingly, the corresponding mixtures with alkanes show relatively low $H^E$ values: 456 J•mol$^{-1}$ for DPA + heptane [19], and 277 J•mol$^{-1}$ for DBA + heptane (at 303.15 K) [20]. The corresponding $V^E$ values are lower than in the case of 2-propanone solutions: 0.268 cm$^3$•mol$^{-1}$ (DPA + heptane) and 0.052 cm$^3$•mol$^{-1}$ (DBA + heptane) [21]. The latter value suggests that structural effects may become important, which is supported by the negative $V^E$ of the DBA + hexane system, $-0.185$ cm$^3$•mol$^{-1}$ [21]. Such effects are also relevant in mixtures including TEA [22], a weakly polar tertiary amine. In solutions with heptane, $H^E = 112$ J•mol$^{-1}$ and $V^E = 0.1255$ cm$^3$•mol$^{-1}$ [23], while in the hexadecane system, $H^E = 322$ J•mol$^{-1}$ [24] $V^E = -0.0979$ cm$^3$•mol$^{-1}$ [25].

We note that for the studied mixtures, $V^E$ is positive. Therefore, the contribution to $V^E$ from the breaking of the interactions between like molecules upon mixing is predominant over the negative contributions from structural effects and interactions between unlike molecules. The existence of such interactions is supported by the $V^E$ decrease observed in 2-propanone mixtures when heptane is replaced by DPA, two solvents of similar size. The strength of the interaction between unlike molecules for the 2-propanone-aniline system has been estimated to be $-30.50$ kJ•mol$^{-1}$ [4]. Moreover, $V^E$, $\alpha_p^E$ and $\kappa_S^E$ are negative for 2-propanone + aniline, + N-metrhylaniline or + pyridine mixtures [4]. In the case of the aniline solution, $V^E = -1.133$ cm$^3$•mol$^{-1}$, $\alpha_p^E = -94.4 \cdot 10^{-6}$ K$^{-1}$ and $\kappa_S^E = -142.5$ TPa$^{-1}$ [4]. This remarks that the amine-ketone interactions are much stronger than in mixtures including DPA, DBA or TEA, which are characterized by positive values of $V^E$, $\alpha_p^E$ and $\kappa_S^E$ (Table 4, Figs, 1-2). Negative $\left(\dfrac{\partial V^E}{\partial T}\right)_P$ values have been interpreted in terms of a decrease in the molar volume of complex formation, which overcompensates for the decrease in the extent of complex formation, and have been encountered, e.g., in amine + trichloromethane mixtures [26,27].

The $V^E$ increase observed when replacing DPA by DBA may be ascribed to the interactions between 2-propanone molecules are broken more easily by DBA, due to its larger aliphatic surface. Moreover, the creation of the amine-ketone interactions is more difficult as the amine group is more sterically hindered in this amine. On the other hand, $V^E$ is higher for the TEA solution than for the DPA mixture. This reveals that the interactions between unlike molecules are more important in the latter system, which is confirmed by the larger $\alpha_p^E$ value encountered for the TEA solution.

The parameter $\chi = \left(\dfrac{u}{u^{id}}\right)^2 - 1$ is widely used to estimate the non-ideality of a system, [28-31] as solutions with strong deviations from the ideal behavior are characterized by high $\chi$ values.

For example, for 2-pyrrolidone mixtures, $\chi$ (methanol) = 0.8 and $\chi$ (ethanol) = 0.35 [29]. In the case of systems containing 2-propanone, $\chi$ (DPA) = $-0.021$ > $\chi$ (DBA) = $-0.041$ > $\chi$ (TEA) = $-0.142$; and $\chi$ (aniline) = 0.463 > $\chi$ (N-methylaniline) = 0.275 > $\chi$ (pyridine) = 0.159. This is in agreement with our previous findings: interactions between unlike molecules become weaker in the sequence DPA > DBA > TEA. Such interactions are much stronger in those systems with aromatic amines.

Finally, we note that the $V^E$ curves are shifted to higher mole fractions of 2-propanone, the smaller component, in solutions with DPA, or DBA (Fig. 1), which is typical of systems where free volume effects are present [18].

## 5. CONCLUSIONS

In this work, we have determined $V^E$, $\kappa_S^E$ and $\alpha_p^E$ for 2-propanone + DPA, + DBA, or + TEA. These excess functions are positive. It is observed that they increase when replacing DPA by DBA or TEA. This may be attributed to interactions between unlike molecules are more important in the DPA solution. The data suggest that the interactions between unlike molecules are weaker than in 2-propanone + aromatic amine mixtures. Free volume effects are present in systems with DPA or DBA, as the $V^E$ curves are shifted towards higher mole fractions of 2-propanone, the smaller component

**ACKNOWLEDGEMENTS**

The authors gratefully acknowledge the financial support received from the Consejería de Educación y Cultura of Junta de Castilla y León, under the Project VA052A09 and from the Ministerio de Educación y Ciencia, under the Project FIS2007-61833. I.A. and V.A. also gratefully acknowledge the grants received from the Junta de Castilla y León.

TABLE 1

Physical properties of pure compounds, 2-propanone, dipropylamine, dibutylamine and triethylamine at temperature $T$: $\rho$, density; $u$, speed of sound; $\alpha_P$, isobaric thermal expansion coefficient; $\kappa_S$, adiabatic compressibility; $\kappa_T$, isothermal compressibility and $C_P$, isobaric heat capacity.

| Property | T/K | 2-propanone | | dipropylamine | | dibutylamine | | triethylamine | |
|---|---|---|---|---|---|---|---|---|---|
| | | Exp | Lit | Exp. | Lit | Exp. | Lit. | Exp. | Lit. |
| $\rho$/g cm$^3$ | 293.15 | 0.790546 | 0.78998[a] | 0.738301 | 0.73720[b] | 0.759591 | 0.762022[c] | 0.727514 | 0.7276[a] |
| | | | | | | | | | 0.72753[d] |
| | 298.15 | 0.784868 | 0.784431[e] | 0.733676 | 0.73336[f] | 0.755549 | 0.75553[f] | 0.722892 | 0.72318[g] |
| | | | 0.78428[h] | | 0.73368[g] | | 0.75570[i] | | 0.72376[j] |
| | | | 0.78457[k] | | 0.73333[l] | | 0.75572[g] | | |
| | | | | | | | 0.75595[j] | | |
| | 303.15 | 0.779169 | 0.77914[h] | 0.729102 | 0.72820[b] | 0.751481 | 0.75194[j] | 0.718332 | 0.71836[d] |
| | | | | | 0.73121[j] | | 0.75248[m] | | |
| | | | | | 0.73019[m] | | | | |
| $u$ /m s$^{-1}$ | 293.15 | 1182.5 | 1192[k] | 1209.1 | | 1261.3 | 1269.47[c] | 1132.7 | |
| | 298.15 | 1160.4 | 1161.72[e] | 1187.8 | 1198[j] | 1241.4 | 1248[j] | 1111.2 | 1123[j] |
| | | | 1154[h] | | | | | | |
| | | | 1160.6[n] | | | | 1246.7[c] | | 1115.1[o] |
| | 303.15 | 1139.2 | 1131.2[h] | 1167.4 | 1174[j] | 1222.6 | 1227[j] | 1091.1 | 1101[j] |
| $\alpha_P$/10$^{-3}$K$^{-1}$ | 298.15 | 1.45 | 1.426[h] | 1.25 | 1.201[j] | 1.07 | 1.059[j] | 1.27 | 1.24[o] |
| $\kappa_S$/Tpa$^{-1}$ | 293.15 | 904.66 | | 926.5 | | 827.6 | 814.31[c] | 1071.29 | |
| | 298.15 | 946.29 | 944.59[e] | 966.1 | 947[j] | 858.8 | 849[j] | 1120.39 | 1113[o] |
| | | | 958[h] | | | | | | |
| | | | 946[n] | | | | | | |
| | 303.15 | 988.80 | 1003[h] | 1006.4 | 992[j] | 890.2 | 883[j] | 1169.44 | 1135[j] |
| $\kappa_T$/Tpa$^{-1}$ | 298.15 | 1317.5 | 1324[a] | 1221.8 | 1183[j] | 1053.4 | 1039[j] | 1432.2 | 1404[o] |
| | | | 1330[h] | | | | | | |
| $C_P$/ J mol$^{-1}$K$^{-1}$ | 298.15 | | 124.9[a] | | 252.84[a] | | 302[j] | | 216.43[p] |

[a] [6]; [b] [32]; [c] [33]; [d] [34]; [e] [17]; [f] [35]; [g] [21]; [h] [36]; [i] [37]; [j] [38]; [k] [39]; [l] [40]; [m] [41]; [n] [42]; [o] [43]; [p] [44]

TABLE 2

Densities, $\rho$, molar excess volumes, $V^E$, and speeds of sound for 2-propanone(1) + amine(2) mixtures at temperature $T$.

| $x_1$ | $\rho$/g cm$^{-3}$ | $V^E$/ cm$^3$ mol$^{-1}$ | $u$ /m s$^{-1}$ | $x_1$ | $\rho$/g cm$^{-3}$ | $V^E$/ cm$^3$ mol$^{-1}$ | $u$ /m s$^{-1}$ |
|---|---|---|---|---|---|---|---|
| colspan="8" | 2-propanone(1) + dipropylamine(2) ; $T$ = 293.15 K |||||||
| 0.0681 | 0.739993 | 0.0474 | 1205.77 | 0.5443 | 0.756750 | 0.2440 | 1184.18 |
| 0.1143 | 0.741208 | 0.0788 | 1203.43 | 0.5928 | 0.759180 | 0.2433 | 1182.35 |
| 0.1667 | 0.742666 | 0.1126 | 1200.78 | 0.6511 | 0.762365 | 0.2354 | 1180.47 |
| 0.2117 | 0.744010 | 0.1360 | 1198.65 | 0.7007 | 0.765284 | 0.2278 | 1179.56 |
| 0.2523 | 0.745280 | 0.1574 | 1196.68 | 0.7489 | 0.768408 | 0.2099 | 1178.75 |
| 0.3075 | 0.747112 | 0.1832 | 1194.10 | 0.7947 | 0.771624 | 0.1879 | 1178.27 |
| 0.3567 | 0.748857 | 0.2034 | 1191.86 | 0.8496 | 0.775792 | 0.1607 | 1178.25 |
| 0.4033 | 0.750616 | 0.2199 | 1189.81 | 0.8996 | 0.780162 | 0.1091 | 1178.88 |
| 0.4448 | 0.752285 | 0.2316 | 1187.98 | 0.9460 | 0.784521 | 0.0643 | 1180.06 |
| 0.4951 | 0.754452 | 0.2409 | 1185.98 | | | | |
| colspan="8" | 2-propanone(1) + dipropylamine(2); $T$ = 298.15 K |||||||
| 0.0541 | 0.735054 | 0.0246 | 1185.12 | 0.5502 | 0.751994 | 0.2458 | 1162.14 |
| 0.1087 | 0.736435 | 0.0637 | 1182.30 | 0.5945 | 0.754194 | 0.2432 | 1160.66 |
| 0.1626 | 0.737883 | 0.1010 | 1179.58 | 0.6466 | 0.756950 | 0.2389 | 1159.08 |
| 0.2081 | 0.739179 | 0.1299 | 1177.24 | 0.6924 | 0.759575 | 0.2307 | 1157.96 |
| 0.2518 | 0.740501 | 0.1547 | 1175.13 | 0.7444 | 0.762840 | 0.2120 | 1156.89 |
| 0.2988 | 0.742014 | 0.1780 | 1172.85 | 0.7963 | 0.766389 | 0.1889 | 1156.28 |
| 0.3473 | 0.743660 | 0.2011 | 1170.58 | 0.8448 | 0.769995 | 0.1624 | 1156.25 |
| 0.3927 | 0.745318 | 0.2177 | 1168.54 | 0.8965 | 0.774358 | 0.1131 | 1156.74 |
| 0.4488 | 0.747521 | 0.2328 | 1166.33 | 0.9443 | 0.778758 | 0.0644 | 1157.98 |
| 0.4918 | 0.749317 | 0.2432 | 1164.45 | | | | |
| colspan="8" | 2-propanone(1) + dipropylamine(2); $T$ = 303.15 K |||||||
| 0.0655 | 0.730643 | 0.0472 | 1163.99 | 0.5527 | 0.746973 | 0.2616 | 1141.10 |
| 0.1125 | 0.731795 | 0.0839 | 1161.46 | 0.5959 | 0.749047 | 0.2618 | 1139.59 |
| 0.1636 | 0.733130 | 0.1193 | 1158.87 | 0.6487 | 0.751779 | 0.2565 | 1137.91 |
| 0.2007 | 0.734156 | 0.1431 | 1156.87 | 0.6959 | 0.754441 | 0.2458 | 1136.70 |
| 0.2594 | 0.735888 | 0.1758 | 1154.06 | 0.7493 | 0.757725 | 0.2265 | 1135.79 |
| 0.3080 | 0.737419 | 0.2007 | 1151.73 | 0.7940 | 0.760769 | 0.1998 | 1135.26 |
| 0.3537 | 0.738962 | 0.2193 | 1149.54 | 0.8513 | 0.765043 | 0.1568 | 1135.20 |

TABLE 2 (continued)

| | | | | | | | |
|---|---|---|---|---|---|---|---|
| 0.4034 | 0.740751 | 0.2364 | 1147.22 | 0.8993 | 0.768986 | 0.1152 | 1135.66 |
| 0.4499 | 0.742530 | 0.2503 | 1145.25 | 0.9486 | 0.773476 | 0.0611 | 1136.92 |
| 0.5001 | 0.744620 | 0.2571 | 1143.12 | | | | |

2-propanone(1) + dibutylamine(2); $T = 293.15$ K

| | | | | | | | |
|---|---|---|---|---|---|---|---|
| 0.0617 | 0.760123 | 0.0738 | 1256.07 | 0.5423 | 0.767568 | 0.4189 | 1213.62 |
| 0.1067 | 0.760523 | 0.1303 | 1252.21 | 0.5934 | 0.768974 | 0.4178 | 1208.95 |
| 0.1645 | 0.761115 | 0.1936 | 1247.33 | 0.6613 | 0.771131 | 0.4044 | 1202.92 |
| 0.2213 | 0.761798 | 0.2443 | 1242.45 | 0.6995 | 0.772514 | 0.3904 | 1199.57 |
| 0.2624 | 0.762317 | 0.2825 | 1238.78 | 0.7456 | 0.774389 | 0.3650 | 1195.69 |
| 0.3027 | 0.762916 | 0.3085 | 1235.31 | 0.7953 | 0.776805 | 0.3140 | 1191.97 |
| 0.3513 | 0.763667 | 0.3424 | 1231.00 | 0.8542 | 0.779951 | 0.2610 | 1187.73 |
| 0.3892 | 0.764326 | 0.3627 | 1227.60 | 0.9052 | 0.783274 | 0.1896 | 1184.96 |
| 0.4490 | 0.765488 | 0.3873 | 1222.20 | 0.9503 | 0.786761 | 0.1065 | 1183.42 |
| 0.4963 | 0.766488 | 0.4059 | 1217.82 | | | | |

2-propanone(1) + dibutylamine(2); $T = 298.15$ K

| | | | | | | | |
|---|---|---|---|---|---|---|---|
| 0.0579 | 0,756021 | 0.0664 | 1236.50 | 0.5507 | 0.763139 | 0.4294 | 1187.27 |
| 0.1092 | 0.756419 | 0.1353 | 1231.98 | 0.5996 | 0.764405 | 0.4304 | 1182.37 |
| 0.1498 | 0.756795 | 0.1808 | 1228.39 | 0.6520 | 0.765956 | 0.4220 | 1178.06 |
| 0.2099 | 0.757425 | 0.2407 | 1223.10 | 0.6999 | 0.767576 | 0.4046 | 1173.52 |
| 0.2537 | 0.757962 | 0.2754 | 1219.33 | 0.7529 | 0.769637 | 0.3743 | 1169.76 |
| 0.3010 | 0.758582 | 0.3121 | 1215.00 | 0.8005 | 0.771785 | 0.3346 | 1165.81 |
| 0.3563 | 0.759402 | 0.3471 | 1210.01 | 0.8556 | 0.774693 | 0.2731 | 1163.20 |
| 0.4086 | 0.760258 | 0.3768 | 1205.30 | 0.9035 | 0.777804 | 0.1894 | 1161.40 |
| 0.4582 | 0.761161 | 0.4003 | 1200.74 | 0.9520 | 0.781451 | 0.0934 | 1187.27 |
| 0.5016 | 0.762025 | 0.4188 | 1196.47 | | | | |

2-propanone(1) + dibutylamine(2); $T = 303.15$ K

| | | | | | | | |
|---|---|---|---|---|---|---|---|
| 0.0563 | 0.751852 | 0.0733 | 1217.73 | 0.4982 | 0.757338 | 0.4321 | 1177.49 |
| 0.1072 | 0.752233 | 0.1369 | 1213.28 | 0.5486 | 0.758426 | 0.4398 | 1172.42 |
| 0.1559 | 0.752644 | 0.1926 | 1208.96 | 0.6004 | 0.759686 | 0.4412 | 1167.37 |
| 0.2030 | 0.753101 | 0.2392 | 1205.08 | 0.6503 | 0.761095 | 0.4303 | 1162.88 |
| 0.2398 | 0.753461 | 0.2789 | 1201.43 | 0.7509 | 0.764591 | 0.3779 | 1153.57 |
| 0.2977 | 0.754122 | 0.3304 | 1196.07 | 0.8002 | 0.766620 | 0.3462 | 1149.45 |
| 0.3441 | 0.754926 | 0.3593 | 1191.88 | 0.8488 | 0.769033 | 0.2919 | 1145.76 |

TABLE 2 (continued)

| | | | | | | | |
|---|---|---|---|---|---|---|---|
| 0.3955 | 0.755499 | 0.3934 | 1187.01 | 0.9066 | 0.772605 | 0.1903 | 1142.50 |
| 0.4560 | 0.756564 | 0.4134 | 1181.34 | 0.9518 | 0.775714 | 0.1150 | 1140.34 |

2-propanone(1) + triethylamine(2); $T = 293.15$ K

| | | | | | | | |
|---|---|---|---|---|---|---|---|
| 0.0599 | 0.729078 | 0.0897 | 1131.39 | 0.5516 | 0.749498 | 0.3738 | 1136.76 |
| 0.1079 | 0.730405 | 0.1587 | 1130.60 | 0.6046 | 0.752888 | 0.3597 | 1139.29 |
| 0.1594 | 0.731971 | 0.2205 | 1130.06 | 0.6537 | 0.756171 | 0.3365 | 1142.21 |
| 0.1894 | 0.732974 | 0.2456 | 1129.92 | 0.7021 | 0.759717 | 0.3141 | 1145.53 |
| 0.2528 | 0.735234 | 0.2962 | 1130.08 | 0.7509 | 0.763637 | 0.2764 | 1149.48 |
| 0.2930 | 0.736792 | 0.3209 | 1130.26 | 0.8040 | 0.768298 | 0.2372 | 1154.65 |
| 0.3369 | 0.738574 | 0.3473 | 1130.75 | 0.8480 | 0.772464 | 0.2003 | 1159.45 |
| 0.4069 | 0.741807 | 0.3692 | 1132.00 | 0.8919 | 0.777086 | 0.1417 | 1164.97 |
| 0.4498 | 0.743891 | 0.3746 | 1133.27 | 0.9515 | 0.784093 | 0.0571 | 1174.03 |
| 0.4940 | 0.746287 | 0.3756 | 1134.45 | | | | |

2-propanone(1) + triethylamine(2); $T = 298.15$ K

| | | | | | | | |
|---|---|---|---|---|---|---|---|
| 0.0611 | 0.724409 | 0.1011 | 1109.75 | 0.5546 | 0.744523 | 0.4018 | 1114.46 |
| 0.1065 | 0.725617 | 0.1718 | 1108.88 | 0.6058 | 0.747700 | 0.3859 | 1116.97 |
| 0.1544 | 0.727020 | 0.2342 | 1108.18 | 0.6549 | 0.751001 | 0.3644 | 1119.87 |
| 0.1958 | 0.728349 | 0.2770 | 1108.09 | 0.7012 | 0.754363 | 0.3391 | 1123.07 |
| 0.2462 | 0.730099 | 0.3200 | 1107.83 | 0.7509 | 0.758296 | 0.3048 | 1127.06 |
| 0.2986 | 0.732114 | 0.3515 | 1107.99 | 0.8018 | 0.762708 | 0.2600 | 1131.90 |
| 0.3582 | 0.734593 | 0.3762 | 1109.00 | 0.8530 | 0.767596 | 0.2054 | 1137.79 |
| 0.3978 | 0.736340 | 0.3923 | 1109.54 | 0.8887 | 0.771293 | 0.1620 | 1142.34 |
| 0.4511 | 0.738891 | 0.4044 | 1110.91 | 0.9501 | 0.778305 | 0.0766 | 1151.43 |
| 0.5024 | 0.741579 | 0.4053 | 1112.40 | | | | |

2-propanone(1) + triethylamine(2); $T = 303.15$ K

| | | | | | | | |
|---|---|---|---|---|---|---|---|
| 0.0707 | 0.719955 | 0.1358 | 1089.13 | 0.5499 | 0.739105 | 0.4327 | 1093.25 |
| 0.1190 | 0.721258 | 0.2063 | 1088.19 | 0.6036 | 0.742346 | 0.4179 | 1095.65 |
| 0.1589 | 0.722370 | 0.2659 | 1087.63 | 0.6534 | 0.745607 | 0.3979 | 1098.71 |
| 0.2188 | 0.724294 | 0.3262 | 1087.30 | 0.7023 | 0.749124 | 0.3670 | 1101.91 |
| 0.2986 | 0.727146 | 0.3898 | 1087.20 | 0.7443 | 0.752427 | 0.3310 | 1105.47 |
| 0.3474 | 0.729113 | 0.4133 | 1087.80 | 0.7964 | 0.756891 | 0.2770 | 1110.38 |
| 0.4004 | 0.731418 | 0.4315 | 1088.59 | 0.8481 | 0.761680 | 0.2241 | 1115.99 |
| 0.4514 | 0.733852 | 0.4372 | 1089.96 | 0.9028 | 0.767332 | 0.1532 | 1123.08 |
| 0.5221 | 0.737827 | 0.4378 | 1092.43 | 0.9496 | 0.772831 | 0.0692 | 1130.64 |

TABLE 3

Excess functions at 298.15 K for $\kappa_S$, adiabatic compressibility and $\alpha_P$, isobaric thermal expansion coefficient of 2-propanone(1) + amine(2) mixtures.

| $x_1$ | $\kappa_S^E$ /TPa$^{-1}$ | $\alpha_P^E$ /10$^{-6}$·K$^{-1}$ |
|---|---|---|
| \multicolumn{3}{c}{2-propanone(1) + dipropylamine(2)} | | |
| 0.0541 | 2.58  | 0.52 |
| 0.1087 | 5.47  | 2.12 |
| 0.1626 | 8.17  | 4.31 |
| 0.2081 | 10.47 | 6.42 |
| 0.2518 | 12.42 | 8.52 |
| 0.2988 | 14.45 | 10.70 |
| 0.3473 | 16.38 | 12.75 |
| 0.3927 | 17.96 | 14.32 |
| 0.4488 | 19.28 | 15.68 |
| 0.4918 | 20.56 | 16.23 |
| 0.5502 | 21.73 | 16.11 |
| 0.5945 | 22.07 | 15.38 |
| 0.6466 | 22.12 | 13.90 |
| 0.6924 | 21.63 | 12.05 |
| 0.7444 | 20.60 | 9.42 |
| 0.7963 | 18.75 | 6.53 |
| 0.8448 | 16.05 | 3.88 |
| 0.8965 | 12.07 | 1.26 |
| 0.9443 | 7.14  | −0.21 |
| \multicolumn{3}{c}{2-propanone(1) + dibutylamine(2)} | | |
| 0.0579 | 3.23  | 2.50 |
| 0.1092 | 6.27  | 4.53 |
| 0.1498 | 8.58  | 5.95 |
| 0.2099 | 11.79 | 7.87 |
| 0.2537 | 13.82 | 9.14 |
| 0.3010 | 16.25 | 10.49 |
| 0.3563 | 18.76 | 12.01 |
| 0.4086 | 20.89 | 13.45 |
| 0.4582 | 22.80 | 14.82 |
| 0.5016 | 24.71 | 16.01 |
| 0.5507 | 26.27 | 17.26 |

Table 3 (continued)

| | | |
|---|---|---|
| 0.5996 | 27.39 | 18.35 |
| 0.6520 | 28.18 | 19.28 |
| 0.6999 | 28.22 | 19.78 |
| 0.7529 | 27.37 | 19.76 |
| 0.8005 | 25.54 | 18.98 |
| 0.8556 | 22.06 | 16.84 |
| 0.9035 | 16.80 | 13.22 |
| 0.9520 | 9.43 | 7.68 |
| 2-propanone(1) + triethylamine(2) | | |
| 0.0611 | 6.28 | 17.68 |
| 0.1065 | 10.68 | 26.81 |
| 0.1544 | 14.92 | 33.78 |
| 0.1958 | 17.54 | 38.21 |
| 0.2462 | 21.15 | 42.31 |
| 0.2986 | 24.09 | 45.59 |
| 0.3582 | 25.97 | 48.73 |
| 0.3978 | 27.63 | 50.61 |
| 0.4511 | 28.76 | 52.82 |
| 0.5024 | 29.69 | 54.44 |
| 0.5546 | 29.88 | 55.38 |
| 0.6058 | 29.33 | 55.16 |
| 0.6549 | 28.21 | 53.61 |
| 0.7012 | 26.64 | 50.68 |
| 0.7509 | 24.35 | 45.69 |
| 0.8018 | 21.17 | 38.51 |
| 0.8530 | 16.78 | 29.34 |
| 0.8887 | 13.41 | 22.09 |
| 0.9501 | 6.53 | 9.12 |

TABLE 4

Coefficients $A_i$ and standard deviations, $\sigma(F^E)$ (eq. 7) for representation of the $F^{E,a}$ property at 298.15 K for 2-propanone(1) + aromatic amine(2) systems by eq. 6

| System[b] | T/K | Property $F^E$ | $A_0$ | $A_1$ | $A_2$ | $A_3$ | $A_4$ | $\sigma(F^E)$ |
|---|---|---|---|---|---|---|---|---|
| 2-propanone + dipropylamine | 293.15 | $V^E$ | 0.960 | 0.286 | 0.077 | | | 0.003 |
| | 298.15 | $V^E$ | 0.971 | 0.252 | −0.02 | 0.20 | | 0.003 |
| | | $\kappa_S^E$ | 83.13 | 40.8 | 16.4 | 9.3 | | 0.11 |
| | | $\alpha_P^E$ | 64.92 | 8.9 | −69.6 | −20.9 | −9.8 | 0.04 |
| | 303.15 | $V^E$ | 1.043 | 0.279 | | | | 0.002 |
| 2-propanone + dibutylamine | 293.15 | $V^E$ | 1.628 | 0.488 | 0.20 | | | 0.003 |
| | 298.15 | $V^E$ | 1.670 | 0.549 | 0.21 | | | 0.005 |
| | | $\kappa_S^E$ | 98.19 | 69.8 | 46.5 | 15.4 | | 0.12 |
| | | $\alpha_P^E$ | 63.62 | 52.6 | 53.9 | 19.5 | | 0.05 |
| | 303.15 | $V^E$ | 1.727 | 0.529 | 0.24 | | | 0.005 |
| 2-propanone + triethylamine | 293.15 | $V^E$ | 1.511 | −0.085 | 0.09 | | | 0.004 |
| | 298.15 | $V^E$ | 1.622 | −0.089 | 0.174 | | | 0.003 |
| | | $\kappa_S^E$ | 118.6 | 16.4 | 10.4 | | | 0.2 |
| | | $\alpha_P^E$ | 218.9 | 54 | 59 | −148 | | 0.3 |
| | 303.15 | $V^E$ | 1.767 | −0.10 | 0.16 | −0.18 | | 0.004 |

[a] $F^E = V^E$, units: cm$^3$ mol$^{-1}$; $F^E = \kappa_S^E$, units: TPa$^{-1}$; $F^E = \alpha_P^E$, units: 10$^{-6}$ K$^{-1}$

**CAPTION TO FIGURES**

Fig. 1. $V^E$ for the 2-propanone(1) + amine(2) systems at atmospheric pressure and 298.15 K. Full symbols (this work): (●), dipropylamiane; (■), dibutylamine; (▲), triethylamine. Solid lines, calculations with eq. (6) using the coefficients from Table 4.

Fig. 2. $\kappa_S^E$ for the for the 2-propanone(1) + amine(2) systems at atmospheric pressure and 298.15 K. Full symbols (this work): (●), dipropylamiane; (■), dibutylamine; (▲), triethylamine. Solid lines, calculations with eq. (6) using the coefficients from Table 4.

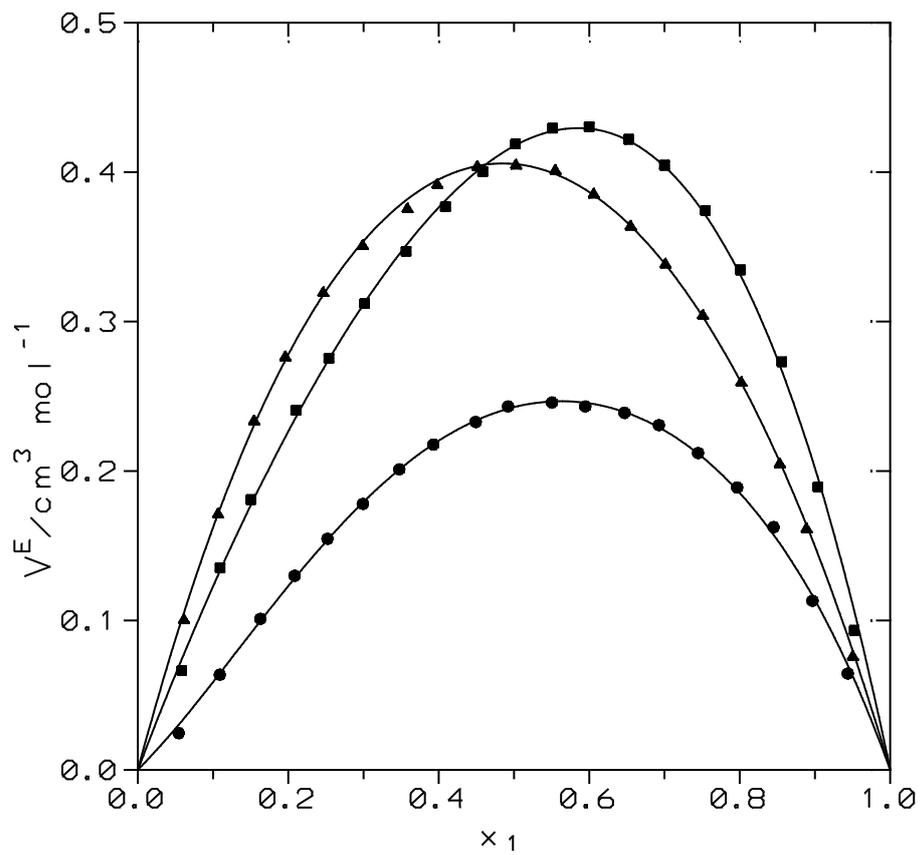

FIG. 1

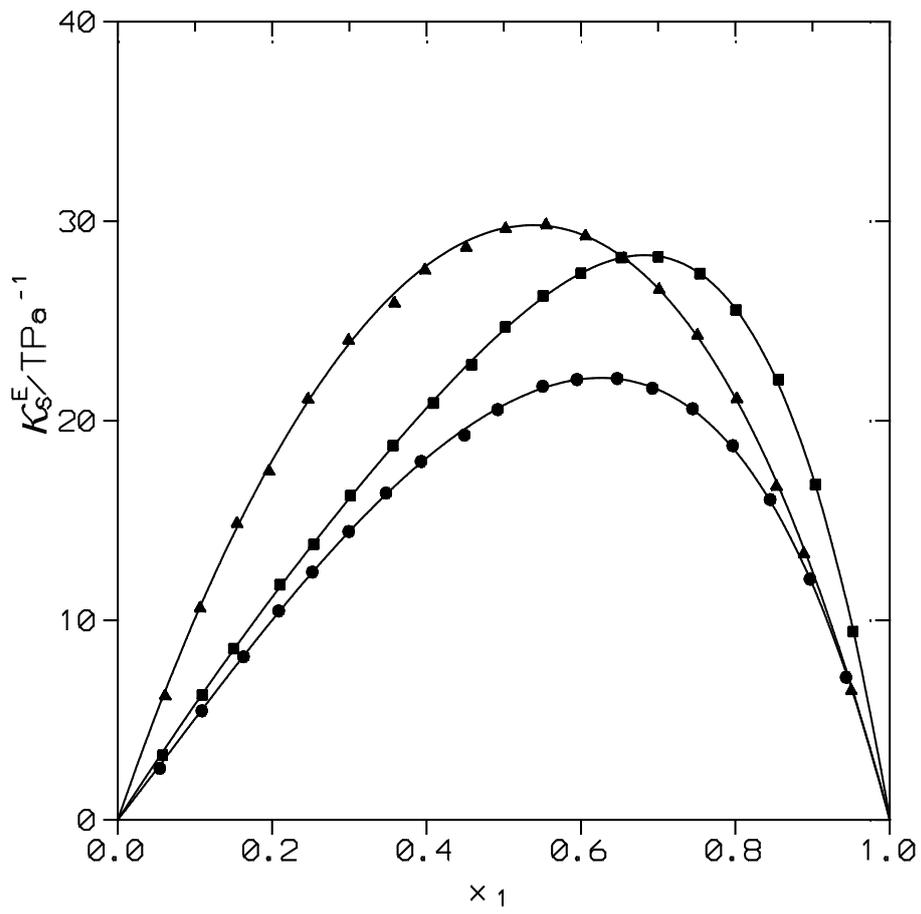

FIG. 2